\documentclass[useAMS,usenatbib,usegraphicx]{mn2e}

\usepackage{graphicx}
\usepackage{times}
\usepackage{amssymb,amsmath,amsfonts,graphicx,aas_macros}
\usepackage{natbib,color}

\usepackage[bookmarks,bookmarksnumbered,colorlinks=true, citecolor=blue, linkcolor=black]{hyperref}

\newcommand{\lcdm}{{$\Lambda$CDM}}
\newcommand{\som}{{\sigma_{\rm T}/m_{\chi}}}  
\newcommand{\cmspg}{{\rm cm^2\, g^{-1}}}
\newcommand{\Msun}{{$\rm M_\odot$}}
\newcommand{\rdisk}{{R_{\rm d}}}
\newcommand{\mdisk}{{M_{\rm d}}}
\newcommand{\zdisk}{{z_{\rm d}}}

\newcommand{\Vmax}{{V_{\rm max}}}
\newcommand{\Vtwo}{{V_{2\,\rm kpc}}}
\newcommand{\kms}{{\rm km\, s^{-1}}}

\begin{document}

\title[SIDM halo diversity]{Spreading out and staying sharp
  - Creating diverse rotation curves via
  baryonic and self-interaction effects}

\author[P. Creasey et al.]{Peter Creasey$^{1}$\thanks{E-mail: \href{mailto:peter.creasey@ucr.edu}{peter.creasey@ucr.edu}}, Omid Sameie$^1$, 
Laura V. Sales$^1$, Hai-Bo Yu$^{1}$\thanks{Hellman Fellow},  Mark Vogelsberger$^2$\thanks{Alfred P. Sloan Fellow}  \newauthor 
and Jes\'us Zavala$^3$  \\
$^1$ Department of Physics and Astronomy, University of California, Riverside, California 92507, USA\\
$^2$ Department of Physics, Kavli Institute for Astrophysics and Space Research, Massachusetts Institute of Technology, Cambridge, MA 02139, USA\\  
$^3$ Center for Astrophysics and Cosmology, Science Institute, University of Iceland, Dunhagi 5, 107 Reykjavik, Iceland }

\maketitle

\begin{abstract} 
  Galactic rotation curves are a fundamental constraint for any 
  cosmological model. We use controlled N-body simulations of galaxies 
  to study the gravitational effect of baryons in a scenario 
  with collisionless cold dark matter (CDM) versus one with a
  self-interacting dark matter (SIDM) component. In particular, we 
  examine the inner profiles of the rotation curves in the velocity 
  range $\Vmax =[30-250] \; \kms$, whose diversity 
  has been found to be greater than predicted by the
  \lcdm\ scenario.  We find that the scatter in the observed rotation curves
  exceeds that predicted by dark matter only 
  mass-concentration relations in either the CDM nor SIDM models.
  Allowing for realistic baryonic content and spatial distributions, however,  
  helps create a
  large variety of rotation curve shapes; which is in
  better agreement with observations in the case of self-interactions
  due to the characteristic cored profiles being more accommodating
  to the slowly rising rotation curves than CDM. We find
  individual fits to model two of the most remarkable outliers of similar $\Vmax$, 
  UGC~5721 and IC~2574, the former a cusp-like rotation curve and the
  latter a seemingly $8$~kpc cored profile. This diversity in
  SIDM arises as permutations of overly concentrated halos with compact 
  baryonic distributions versus underdense halos with
  extended baryonic disks. The SIDM solution is promising
  and its feasibility ultimately depends on the sampling of the
  halo mass-concentration relation and its interplay with the baryonic
  profiles, emphasising the need for a better understanding of the frequency
  of extreme outliers present in current observational samples.
\end{abstract}


\section{Introduction}
\label{sec:intro}
Dark matter makes up $80$\% of the total mass budget in the Universe,
but its exact nature remains doggedly elusive. Without direct
detection of a single dark matter particle we have to infer its
properties from its macroscopic distribution, and any observable that
might hint at its underlying particulate nature is seized upon with
vigour. A plethora of observations, such as large scale
structures, lensing and scaling relations, and early fluctuations on
the Cosmic Microwave Background, agree surprisingly well with
predictions of a rather simple cosmological model with a cosmological
constant, denoted as $\Lambda$, and a collisionless cold dark matter
(CDM) component~\citep[e.g.][]{Ade:2013zuv}. This \lcdm ~paradigm has
become the standard and most widely accepted model today for the
observed Universe.

\lcdm\ makes testable predictions at galactic scales that can be
contrasted with observations. Among the most fundamental of them is
for the structure of dark matter halos, with a density profile
parameterised by a Navarro-Frenk-White (NFW)
profile~\citep{Navarro_1997}, which at the inner radii scales as a power
law $r^{-\alpha}$ with $\alpha\approx1$. Observations, however, of the
rotation curves of spiral galaxies, including dwarf and low surface
brightness galaxies, often exhibit an inner circular velocity of stars and
gas that increases more mildly than expected from a CDM
halo~\citep[e.g.
][]{Flores:1994gz,Moore:1994yx,Persic:1995ru,Kuzio_2008,Oh_2015},
indicating an inner density profile shallower than the NFW cusp, i.e.,
$\alpha<1$. This discrepancy can be further generalised as the mass
deficit problem: the CDM halo contains too much dark matter mass in
the inner regions than inferred from
observations~\citep{MBK_2011,Ferrero_2012, Papastergis_2015, Papastergis2016, Schneider_2016}.

A more intriguing observation is that the galactic rotation curves exhibit a large diversity.
Individual fits to galaxy rotation curves 
span a spectrum from 
cores $\alpha\approx 0$ \citep[e.g.][]{Cote2000,deBlok_2008,deNaray:2009xj} to cusps \citep[e.g.
][]{vandenbosch2000,Swaters_2003, Spekkens2005}, and in the case
of cored profiles, the central densities can differ by factor of $10$
for galaxies inhabiting similar halos~\citep{deNaray:2009xj}.
Recently, \cite{Oman_2015} quantified this diversity in rotation
curves by comparing $\Vtwo$ for a fixed $\Vmax$, where $\Vtwo$ is the
circular velocity at $2~{\rm kpc}$ and $\Vmax$ is the peak circular
velocity. For $\Vmax$ in the range of $50\textup{--}250\;\kms$,
the scatter in $\Vtwo$ is a factor of $3\textup{-}4$ and consequently
the mechanism invoked to generate cored profiles must also accommodate
this large variations in rotation curve shapes.
Notably mass modelling is subject to significant uncertainties,
especially at the scale of low mass dwarfs where pressure support
effects, triaxiality, inclination, hidden bars and other
irregularities hamper the utility of
circular velocity profiles for mass modelling
\citep{Hayashi2004,Rhee2004,Pineda_2016,Read2016},
making it hard to assess the precise spread in the rotation curves. In this paper, we take the result reported in Oman et al. as our reference.

The distribution of rotation curves is clearly too heterogeneous to be 
indexed with a single parameter. For example, several different galaxy formation
models seem able to reproduce the Tully-Fisher \citep{TullyFisher} relation within
\lcdm\ \citep{McCarthy_2012,Brook_2016,
  DiCintio_2016, Sales_2016, Ferrero_2016, Santos-Santos2016,
  Katz:2016hyb}  provided
the stellar feedback populates dark matter halos with the right
stellar mass and size \citep[though see also][]{Pace:2016oim},
yet despite this global consensus the circular velocity profiles
predicted by each model differ in detail, with some cases leading to the
formation of cores \citep[e.g.][]{Governato_2010,
  Pontzen_2012, DiCintio_2014,Read:2015sta, Wetzel_2016} whereas other
simulations preserve the inner dark matter cusps
\citep[e.g.][]{Sawala:2014baa, Vogelsberger_2014,Schaller2015}. The
disagreement may arise from the inclusion of different physical
processes, different numerical implementation or may even vary with
star formation histories~\citep[e.g.][]{Onorbe2015, Tollet2016}.

In particular, the feedback model applied by \citet{Oman_2015} appears
unable to explain the scatter in observations within the $\Lambda$CDM
framework. On the other hand, \citet{Brook2015b} finds good agreement
between simulations and observations when looking at the scatter in
the ratio between the circular velocity at $1$ kpc and $\Vmax$.
Similarly \citet{Read2016} has also found consistency between baryon
induced cores within $\Lambda$CDM and the shape of observed rotation
curves. It is unclear, however, that this baryonic solution would hold if the
size of the cores measured in observations is $\gtrsim 1$~kpc
\citep[see for instance][]{Tollet2016}. In fact, this may represent a
serious limitation of this solution. For example IC~2574
has an inferred cored inner halo which extends for $8 \; \rm kpc$,
well beyond the radius of the stars, with a stellar half-mass radius
of $5\; \rm kpc$. This type of object is a challenge to the hypothesis
of feedback generated cores since, by construction, the extent of the
cores in such scenarios is limited to the region where stars form and deposit their
energy~\citep{DiCintio_2014}.

An alternative solution is to consider cores formed out of
self-interacting dark matter (SIDM), which we explore in this paper.
The SIDM model retains important features of the CDM model
including the distribution of halo concentrations as a function of mass  \citep{Rocha:2012jg}. 
SIDM differs, however, in that scattering between dark matter particles leads to heat transfer and the generation of dense cores in the
inner regions of halos ~\citep{Spergel:1999mh,Firmani:2000ce}.
Recent
high-resolution N-body simulations have shown that the
self-interaction cross-section per unit mass $\som\sim1~\cmspg$
is required to have core densities that are preferred by galaxy
observations
\citep{Vogelsberger_2012,Zavala:2012us,Rocha:2012jg,Elbert:2014bma}.
\cite{Kaplinghat_2016} finds $\som\approx1\textup{--}3~\cmspg$
by directly fitting the rotation curves of dwarf galaxies. On the
other hand, there are various constraints on $\som$, including merging
clusters~\citep{Randall_2008,Kahlhoefer:2013dca,Robertson_2016,Kim:2016ujt},
shapes of elliptical galaxies and galaxy
clusters~\citep{MiraldaEscude_2002,Feng:2009hw,Peter_2013}, core sizes of
clusters~\citep{Yoshida_2000,Rocha:2012jg,Kaplinghat_2016,Elbert_2016},
and survival of dwarf halos from evaporation~\citep{Gnedin_2001}.

Among these constraints the strongest limit is $\som \lesssim 0.1\;\cmspg$ in
galaxy clusters~\citep{Yoshida_2000,Kaplinghat_2016,Elbert_2016},
where the relative dark matter velocity $v\sim1500~{\rm km~s^{-1}}$.
As such, the self-interaction cross-section should have a mild velocity
dependence, which can be naturally realised in a class of hidden
sector dark matter models with a light force
mediator~\citep[e.g.][]{Feng:2009mn,Feng:2009hw,Buckley:2009in,Loeb:2010gj,Tulin:2013teo,Tulin:2012wi,Boddy:2014yra,Boddy:2016bbu}.
A velocity-dependent SIDM model, based a Yukawa potential, has been
implemented in zoom-in N-body
simulations~\citep{Vogelsberger_2012,Vogelsberger:2015gpr}, and 
further generalised in the ETHOS (Effective Theory of
Structure Formation) framework, mapping
underlying particle physics parameters to astrophysical
observables~\citep{Vogelsberger:2015gpr,Cyr-Racine:2015ihg}. 

This approach has several distinct features that help explain the
rotation curve diversity. Firstly, the
scatter in the halo concentration leads to variations in core density
directly~\citep{Kaplinghat_2016}. Secondly, the self-interactions
thermalise the inner halo with its central dark matter density
\emph{dependent} upon the baryonic extent. In dark matter dominated
galaxies, a dense core forms whose density is determined by the
self-interaction cross-section and the halo mass. With a cored profile, the features in
the baryon distribution are more prominently reflected in the rotation
curves. In contrast, in galaxies where the baryon component dominates
the potential, the thermalisation process lead to a denser central
core with sizes influenced by the baryonic scale
radius~\citep{Kaplinghat_2014}. Analytical calculations to address
the diversity problem have been carried out by \citet{Kamada2016} using isothermal
approximations to the effects of SIDM with a baryonic potential in
order to construct fits for a diverse range of individual spiral
galaxies.

In this work we explore the combined effects of SIDM and baryonic
potentials by performing a series of numerical experiments.
Our focus is on global trends but we include a
pair of individual fits to test our results on the strongest outliers
from observations. We sample a realistic range of concentrations and
halo masses taken from cosmological simulations together with observed
trends in the mass-size relation of galaxies. Our numerical approach
and the analytical one presented in~\citet{Kamada2016} complement each
other in addressing the diversity problem in the SIDM model. The
structure of this paper is as follows. In Section~\ref{sec:sims} we
introduce in detail our simulations. Sections.~\ref{sec:V2_Vmax_dm} and
\ref{sec:V2_Vmax_bar} explore the diversity problem for a large sample
of halos with $\Vmax=[30$-$250]$~km~s$^{-1}$ with the latter including the effect
of baryons. In Sec.~\ref{sec:obs_curve_sims} we
confirm that we can find rotation curves that are reasonable matches
to a pair of extreme (one cusp-like and one core-like) observed
rotation curves. In Sec.~\ref{sec:discussion} we summarise and
conclude.

\section{Numerical simulations}\label{sec:sims}

We use a combination of N-body simulations and analytical tools to
study the mass profile of galaxies within the SIDM framework including
the effects driven by their baryons. The main focus of this work is on
disk dominated galaxies, which we model with a disk
component embedded within a massive and more extended dark matter halo
(i.e. no bulge). For the N-body simulations we must produce
discretised initial conditions which we describe for the halo in
Section~\ref{sec:bar_ic} in addition to the baryonic potential in
\ref{sec:diskpot}. We evolve these initial conditions with and without
the effect of a baryonic disk and also with and without
self-interaction terms using the simulation code described in
Section~\ref{sec:nbody}.

\subsection{N-body code}\label{sec:nbody}
We use the {\sc Arepo} code \citet{Springel_2010} with the modifications of 
\citet{Vogelsberger_2012} and \citet{Vogelsberger:2015gpr} to account for dark matter self-interactions.
This methodology uses a Monte-Carlo approach
to model the scattering of dark matter via the probabilistic
scattering of the macroscopic dark matter particles in the simulation,
where the density distribution due to an individual particle is
smoothed over some kernel which extends over the $k$-nearest
neighbours. Such an explicit method requires a time step limit $\Delta
t_{\rm sidm}$ such that only $\lesssim 1$ scattering can occur per
particle per timestep. Notably in the cusp region of an NFW halo (i.e.
$\rho \sim r^{-1}$ as $r\to 0$) as one moves towards the centre one
finds that $\Delta t_{\rm sidm} \propto \frac{1}{\rho \sigma} \sim
r^{\frac{1}{2}}$ (with $\sigma$ the velocity dispersion),
 i.e the centre of the halo is so dynamically cold
and dense that the mean-free path is below the particle separation. To
avoid such small time steps one additionally needs to limit them to a
small fraction of the acceleration time step.

We have modified this code to include a baryonic potential of a
Miyamoto-Nagai (hereafter MN) disk (Section~\ref{sec:diskpot}) and
used a constant self-interaction cross-section for scattering, which
is a good approximation in the dark matter low velocity limit, e.g.,
$v\lesssim 400 \; \kms$ as shown in Fig.~1 of \citet{Kaplinghat_2016}.

In our calculations we assume a Hubble constant
$H_0 = 70\; \kms\,{\rm Mpc}^{-1}$.  Our simulation period is fixed at
$10$ Gyr, slightly shorter than the Hubble time
($H_0^{-1} \approx 13.96$~Gyr) in order to account, in an approximate
way, for the assembly time of such galactic disks. It should be
emphasised that isolated simulations with a static disk potential is
well-motivated in the SIDM model; for $\som\gtrsim1~{\rm cm^2/g}$,
dark matter self-interactions occur multiple times in the inner halo
over the age of galaxies, which drive thermalisation of the inner halo
in the presence of the stellar disk, and as such the final inner SIDM
halo profile is more robust to changes in the formation history (given
a known baryon distribution) than that of CDM.

\subsection{Isolated halo initial conditions}\label{sec:bar_ic}

For each of our galaxies, we generate a large set of compound galaxies
consisting of an exponential disk embedded in a dark matter halo using
the code {\sc MakeDisk} \citep{Hernquist_1993, Springel_2005}. The
systems are set up in near equilibrium by requiring a joint solution
to the combined phase space of both disk and halo component.  In
particular, the density profile of the halo is set up with a
\citet{Hernquist_1990} density profile:
\begin{equation}
\rho_{\rm dm}(r) = \frac{M_{\rm dm} r_{\rm hq}}{2\pi r \left(r+r_{\rm hq}\right)^3} \, ,
\end{equation}
with $r_{\rm hq}$ the scale radius of the halo. \citet{Hernquist_1990} profiles
show a $r^{-1}$ slope in the inner density profile and $r^{-4}$ in the
outskirts, in close agreement with the inner/outer slopes of $-1$ and
$-3$, respectively, of the cosmologically inspired NFW
profiles \citep{Navarro_1997}. The Hernquist profile has the extra
advantage of having an analytic expression for its distribution
function, which facilitates the process of setting up the initial
conditions. 

Hernquist halos can be ``scaled'' to match a given NFW
profile quite closely and we therefore report our results in terms of
virial mass\footnote{Virial quantities are defined at the virial
  radius $r_{200}$, where the enclosed density is $200$ times the critical
  density of the Universe. } $M_{200}$ and concentration parameter $c$
corresponding to NFW halos. Finding the equivalent 
halo (by matching the dark matter mass within $r_{200}$ and the 
asymptotic density profile as $r\to0$) gives the relation between parameters 
\begin{eqnarray}
M_{\rm dm} &=& M_{200} - \mdisk \\
r_{\rm hq} &=& \left(\frac{ G M_{200}}{100 H_0^2}\right)^{1/3} c^{-1} \sqrt{2 \ln (1 + c) - \frac{2c}{1+c}} \label{eq:rhq}
\end{eqnarray}
dependent upon the baryonic component $\mdisk$ and the redshift
zero Hubble constant $H_0$.

The matching of the internal regions of our Hernquist profile to the
desired NFW is less accurate as we move close to the scale radius of
the NFW $r_{\rm s}$, with the density of a Hernquist halo falling off more
steeply beyond that point. In terms of the velocity dispersion
profile, the Hernquist halo has a peak velocity dispersion shell that 
is more compact and this velocity dispersion is lower than the NFW value.
This can affect the heat exchange, which stops once the
isothermal core has been established. As a consequence of the lower
velocity dispersion peak, Hernquist profiles can be more prone to the
so called ``core collapse'' phenomena, or the runaway collapse in the
inner regions that ultimately reverses the process of mass scattering
away from the core and condenses instead more mass in the inner
regions. This process would be exacerbated by the presence of baryons
and as $\som$ increases.

We have extensively tested this scenario and we include some comparisons
in Appendix~\ref{sec:NFW_vs_HQ}. We find 
that it makes a negligible difference in
most of our simulations except for extreme cases with concentrated
halos, however as we shall see later it is the \emph{low} concentration
halos where most of the interesting discrepancy between CDM and SIDM predictions lie, i.e. 
these halos are expected to host the low central (dark matter) 
density galaxies that are most challenging to CDM. As such our approach of
using Hernquist halos is conservative, as the effect will be small and any
deviations will cause us to underestimate our conclusions.

For the simulations showed in Sec.~\ref{sec:results}, we set up our
halos using $n_p=10^5$ particles to model the dark matter
distribution, which results in a spread for the mass per particle
${\rm m_p}=8.97\times 10^4$--$2.66\times 10^7 \; \rm M_\odot$ according to the mass of the simulated halo.
We use a gravitational softening length $\epsilon=50$~pc, 
and we have tested for numerical convergence using a
halo with $M_{200}=1.1\times 10^{11} \; \rm M_\odot$ and increasing/decreasing the
particle number by a factor $10$ with respect to our 
fiducial runs and find agreement within $10\%$ in the
(dark matter only) circular velocity profiles (and by proxy mass) for all radii outside $1.0$~kpc. 

\subsection{Disk potential}\label{sec:diskpot}
The baryonic component of our galaxies is modelled by a fixed disk
potential, which is not only computationally cheaper, but also allows
us to maintain full control on the exact distribution of the baryons.
This is particularly important for dwarf galaxies which undergo hundreds
of orbits over a Hubble time and are thus particularly sensitive to 
(unphysical) discretisation-induced instabilities. 
We therefore strip the disk particles created
in our initial conditions, and use instead an extra component in the
gravitational force that accounts for the removed disk particles. 
This approach is not fully self-consistent since it
ignores the back-reaction of the halo onto the baryons, but it provides
a useful tool to explore the evolution of halo potentials in SIDM in a
set of controlled experiments where the {\it final} baryonic
distribution is known. 

We implemented a static potential in {\sc Arepo} following a
\citet{Miyamoto_Nagai_1975} disk:
\begin{equation}
\Phi_{\rm MN}(R,z) = \frac{-G\mdisk}{\sqrt{R^2+\left(\rdisk+\sqrt{\zdisk^2+z^2}\right)^2}} \, ,
\end{equation}
for a disk of mass $\mdisk$ and scale radius $\rdisk$ and we keep the disk relatively
thick in all cases by setting $\zdisk/\rdisk=0.3$ (appropriate to model
dwarf irregular galaxies). Note that the half-mass radius of the flat 
($\zdisk=0$) MN disk is $\sqrt{3}\rdisk$ which is very close 
to the ratio of half-mass to scale radius of an exponential disk of 
$1.678$ and so unless ones disk is extremely thick, the scale radius 
of an exponential fit will be within a few percent\footnote{The
  differences are much more pronounced in the tails of the
  distribution, indeed one cannot choose to correspond a flat
  exponential to a MN density profile via for example the
  mass weighted root-mean square radii since for the latter that is a
  divergent quantity.}. From here on we will use $\rdisk$ interchangeably
to refer to either the scale length of our baryonic disks in simulations
(the MN scale radius) or the exponential disks associated with observed
late-type dwarf galaxies.

Notice that the initial conditions are created assuming an exponential
profile while the fixed potential uses an MN model for the
disk. We have tested that this small inconsistency does not affect our
results.  Due to the diffusive nature of SIDM the resulting deviations
due to non-equilibria are of the order of a few percent and quickly
damped. Alternative approaches include that of \citet{Elbert_2016}
where the MN disk is `grown' adiabatically (compared to the orbital
timescale) from an initially halo-only setup. We have tested the extreme case 
of inserting the fixed potential instantaneously and find that after 10~Gyr 
evolution in SIDM at 2~kpc this makes a relative difference in the $\Vtwo$ 
of $-0.2\;\kms$ or $0.6\%$. Our method, being closer to
equilibrium, is expected to be lower than these bounds. 

We run the simulations for $10$ Gyr and compute the circular velocity
profiles. For the CDM cases this is in the most part a straightforward
numerical test since the initial conditions are created in equilibrium
and we do not see significant evolution of the mass distribution with
time. For the most extended disks ($\rdisk=6\; \rm kpc$), however, the
centre-of-mass is not well `tied' to the centre of the (shallow)
baryonic potential, and the centre-of-mass can `drift' on the order of
a kpc over 10~Gyr (i.e. an average of $0.1\;\kms$) from the centre of the
baryonic disk potential, making the assessment of the mass
distribution within 2~kpc problematic.  This evolution is expected due
to discretisation of ICs and asymmetry in tree-based gravitational
methods, and is relatively hard to suppress even with a tight timestep
criteria. This process can of course also occur to the SIDM halos,
although it is somewhat suppressed since their profiles have less
compact centres.  Since evolving CDM halos in extreme cases is
somewhat tangential to the purpose of this paper, however, we have
tested the CDM evolution on the more compact ($\rdisk=0.5$ and
$1.5\; \rm kpc$) halos, whose $\Vtwo$ evolve by an average of $-7.7\%$
over $10$~Gyr. For the $\rdisk=6\; \rm kpc$ CDM halos we will use the
initial conditions, as we believe those to be more accurate.

\section{Results}
\label{sec:results}

\subsection{Circular velocity profiles in dark matter only halos}
\label{sec:V2_Vmax_dm}

To illustrate the effect of dark matter self-interactions on the density profile of a dwarf halo without the influence of baryons, we use the analytical Jeans method proposed in~\citep{Kaplinghat_2014, Kaplinghat_2016} and tested against simulations in~\citep{Rocha:2012jg,Elbert:2014bma}. We have further checked that its accuracy is within $10\textup{--}20\%$ compared to the results from our N-body code~\citep{Vogelsberger_2012}. The details of the formalism can be found in Appendix~\ref{sec:jeans}.

\begin{table*}
\begin{center}
\begin{tabular}{ccccccc}
\hline
$V_{200}$ & $M_{200}$ &  concentration & $r_{\rm s}$ (kpc)  & $\rho_{\rm s}$ ($10^6 \; \rm M_\odot \; kpc^{-3}$)$^\star$ & $\rm \mdisk$ & ${\rm med.} \; \rdisk$\\
($\kms$) & ($\rm M_\odot$) & ($-2\sigma$,median,$+2\sigma$) & $-2\sigma$,med.,$+2\sigma$ & $-2\sigma$,med.,$+2\sigma$ & ($\rm M_\odot $) & ($\rm kpc^{\star \star}$)\\
\hline
$15$ & $1.12\times 10^{9}$ & $9.99$, $15.60$, $24.38$ & $2.15$, $1.37$, $0.88$ & $6.07$, $18.42$, $57.80$ & $\dagger$ & $\dagger$ \\
$30$ & $8.97\times 10^{9}$ & $8.32$, $12.99$, $20.30$ & $5.15$, $3.30$, $2.11$ & $3.89$, $11.63$, $36.04$ & $  1.71\times 10^{8}$ & ($1.93$) \\
$50$ & $4.15\times 10^{10}$ & $7.27$, $11.36$, $17.74$ & $9.83$, $6.29$, $4.03$ & $2.82$, $8.32$, $25.52$ & $  8.84\times 10^{8}$ & ($2.53$) \\
$70$ & $1.14\times 10^{11}$ & $6.65$, $10.39$, $16.23$ & $15.04$, $9.62$, $6.16$ & $2.29$, $6.69$, $20.36$ & $  2.79\times 10^{9}$ & ($3.03$) \\
$100$ & $3.32\times 10^{11}$ & $6.05$, $9.46$, $14.78$ & $23.60$, $15.11$, $9.67$ & $1.84$, $5.31$, $16.05$ & $  1.09\times 10^{10}$ & ($3.60$) \\
$150$ & $1.12\times 10^{12}$ & $5.44$, $8.50$, $13.28$ & $39.41$, $25.22$, $16.14$ & $1.43$, $4.10$, $12.27$ & $  3.72\times 10^{10}$ & ($4.09$) \\
$200$ & $2.66\times 10^{12}$ & $5.04$, $7.88$, $12.30$ & $56.69$, $36.28$, $23.22$ & $1.20$, $3.42$, $10.15$ & $  6.22\times 10^{10}$ & ($4.53$) \\
\hline
\end{tabular}
\end{center}
\caption{Parameters for the halos used in our diversity analysis. $V_{200}$ and $M_{200}$, concentrations with $\pm 2 \sigma$ relative to the median relation from \citet{Ludlow_2014}, and in NFW parameters we give the scale radius $r_{\rm s}$ and density $\rho_{\rm s}$ for this range of concentrations. For the simulations where we include a baryonic disk, $\rm M_d$ refers to the total baryonic disk mass.\newline
  ${}^\star$ NFW density $\rho_{\rm s}$ is computed in the dark matter-only halo case. For the simulations with baryonic disks we fix $M_{200}$ but add a baryonic component, and thus $\rho_{\rm s}$ is multiplied by the dark fraction $M_{\rm dm}/M_{200}$.\newline
  ${}^{\star\star}$ We quote median baryonic disk radii only for reference, since we attempt to span the distribution with $0.5$-$6\;\rm kpc$ as described in the main text.\newline
${}^\dagger$ This halo was simulated in DM-only.}
\label{table:bar_param}
\end{table*}

Our suite of halos aims to span the `dwarf' galaxy distribution with 7 halos of $V_{200}$ from $15$-$200~\kms$, the upper limit being near Milky Way-like (where velocity dependent corrections may play some role) down to dwarfs so baryon-poor that we expect negligible deviations from the DM-only results. To encompass the diversity in concentrations we sample $\pm 2\sigma$ deviations from the cosmological
mass-concentration relation presented in \citet{Ludlow_2014}, whose median
is well approximated by the linear fit
\begin{equation}\label{eq:linear_ludlow}
c = 10.51 \times \left(\frac{M_{200}}{10^{11} \; \rm M_\odot}\right)^{-0.088} \, ,
\end{equation}
and at each halo mass the distribution is approximately log-normal where $1\sigma$ corresponds to a ratio of $\approx 1.25$. The corresponding halo mass ($M_{200}$) and NFW $r_{\rm s}$ and $\rho_{\rm s}$ are given for the 
$70\; \kms\, \rm Mpc^{-1}$ cosmology in Table~\ref{table:bar_param}.

\begin{figure*}
\includegraphics[width=\columnwidth]{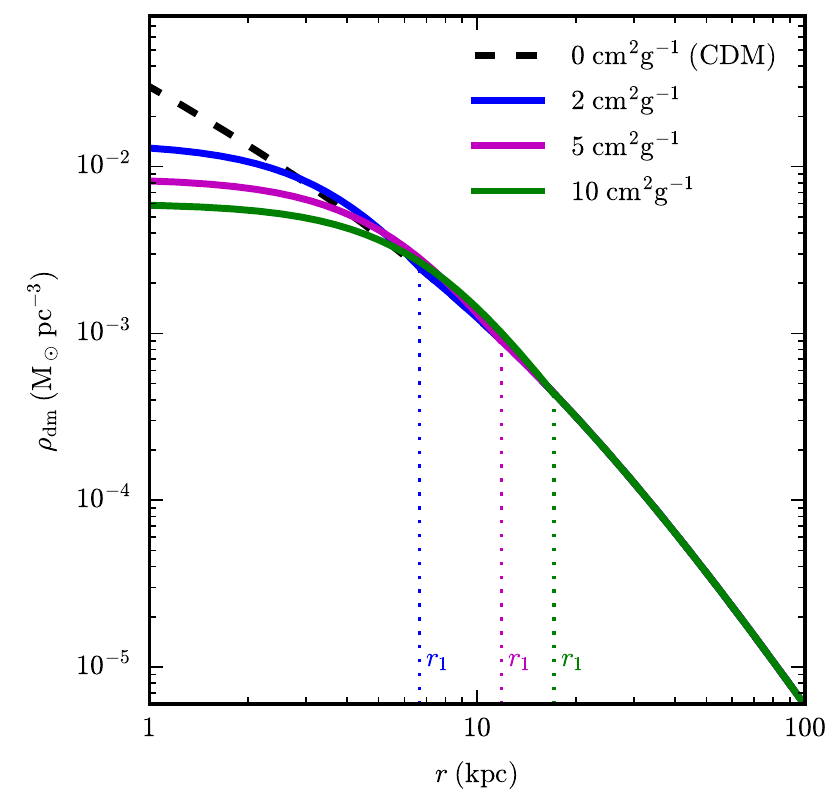}
\includegraphics[width=\columnwidth]{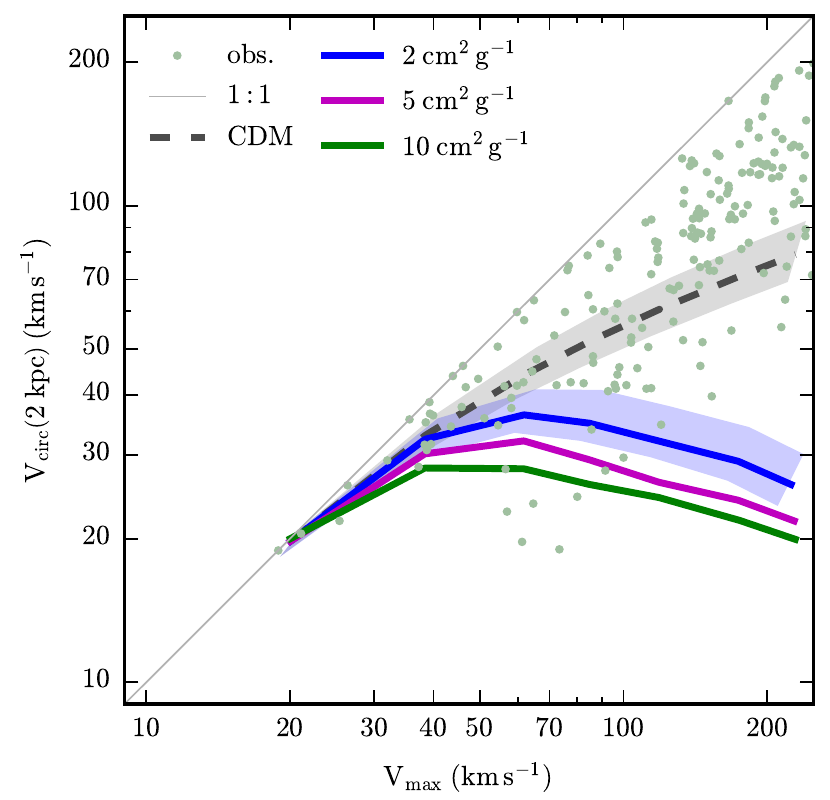}
\caption{\emph{Left panel:} core formation in a dark matter only halo (NFW) with different
  $\som$. Dark matter density profile evolution for SIDM over $10$~Gyr
  in \emph{blue}, \emph{magenta} and \emph{green} for $\som=2$, $5$
  and $10\; \cmspg$ respectively, with CDM ($\som=0\; \cmspg$) in
  \emph{black dashed}. \emph{Vertical dotted lines} indicate $r_1$ (the radius of unit scattering). \emph{Right panel:} effects of SIDM on the $\Vtwo$ vs $\Vmax$
  relation compared to CDM for dark matter only
  halos. \emph{Black line} is the CDM case, \emph{blue} the
  $\som=2\; \cmspg$ where the shaded regions denote $\pm 1 \, \sigma$
  scatter (due to halo concentrations; we have interpolated the values from the $\pm 2\sigma$ runs in Table~\ref{table:bar_param}). \emph{Magenta} and
  \emph{green} denote $\som=5\; \cmspg$ and $10\; \cmspg$
  respectively. \emph{Green points} indicate observed values collated in \citet{Oman_2015}~Table~1, from which we also include the mass deficits, $5\times 10^8$ and $10^9$\Msun and their effect on $\Vtwo$ in 
\emph{blue shaded regions}.}
\label{fig:dm_only}
\end{figure*}

The left panel
of Fig.~\ref{fig:dm_only} shows the density profiles of a halo with
$V_{200}=70 \; \kms$
($M_{200}=1.1\times 10^{11} \; \rm M_\odot$) and a concentration
$c=6.65$ after $10$~Gyr evolution with CDM ($\som=0\; \cmspg$, black
dashed line) vs. SIDM with $\som=2$, $5$ and $10\; \cmspg$ (see solid
lines according to labels). 
The radius of unit scattering, $r_1$ (see also Appendix~\ref{sec:jeans}) outside of which the halo is undisturbed is marked for the three cases. This radius grows with cross-section, and in the interior we see all
three cross-sections produce cores for this halo that extend beyond $2$~kpc.

Fig.~\ref{fig:dm_only} (right panel) shows the circular velocity of the SIDM halo profiles evaluated at $2~{\rm kpc}$ as a function of $V_{\rm max}$. For the comparison, we also plot the corresponding CDM one with the
mass-concentration relation from~\citet{Ludlow_2014} and the range spanned by a compilation of the observed rotation curves in galaxies taken from~\citet{Oman_2015}. It is clear that $\Vtwo$ increases steadily with
$V_{\rm max}$ in the case with CDM, although the relation stays below the $1:1$ line. In contrast, SIDM predicts a much shallower relation. $\Vtwo$ is almost independent of $\Vmax$ in the range explored, and the median $\Vtwo$
value at a given $\Vmax$ depends mildly on cross-section.  For low mass halos (low circular velocity),
$\Vmax$ tends to converge to near $\Vtwo$, since the objects are 
small and the core size becomes much less than $2~{\rm kpc}$ and $\Vmax$ is measured near $2\, \rm kpc$. It is interesting to note this is a common feature in CDM as well as SIDM halos. Shaded regions indicate
$\pm 1 \sigma$ in concentration (we interpolate from the $\pm 2 \sigma$).

The range of $\Vtwo$ and $\Vmax$ statistics of the galaxy rotation curve distribution is shown in Fig.~\ref{fig:dm_only} \citep[taken from][]{Oman_2015}.
The scatter found in the data is significantly
larger than derived purely from that of concentration variations (in either CDM or SIDM). 
A large fraction of the observed $\Vtwo$ scatter
below the relation expected for CDM, suggesting that such objects have
a lower dark matter density in the inner regions than predicted by the
model. Instead, self-interactions seem to have the opposite
behaviour, providing a better  match to these low density objects but
predicting too low a $\Vtwo$ to explain most of the data. These
calculations, however, ignore the contribution of the baryons, and as
we argue below, the mass and radial extent of the gas and stars in
galaxies may significantly change this prediction to bring SIDM in
closer agreement with observations.

\begin{figure*}
  \includegraphics[width=2\columnwidth]{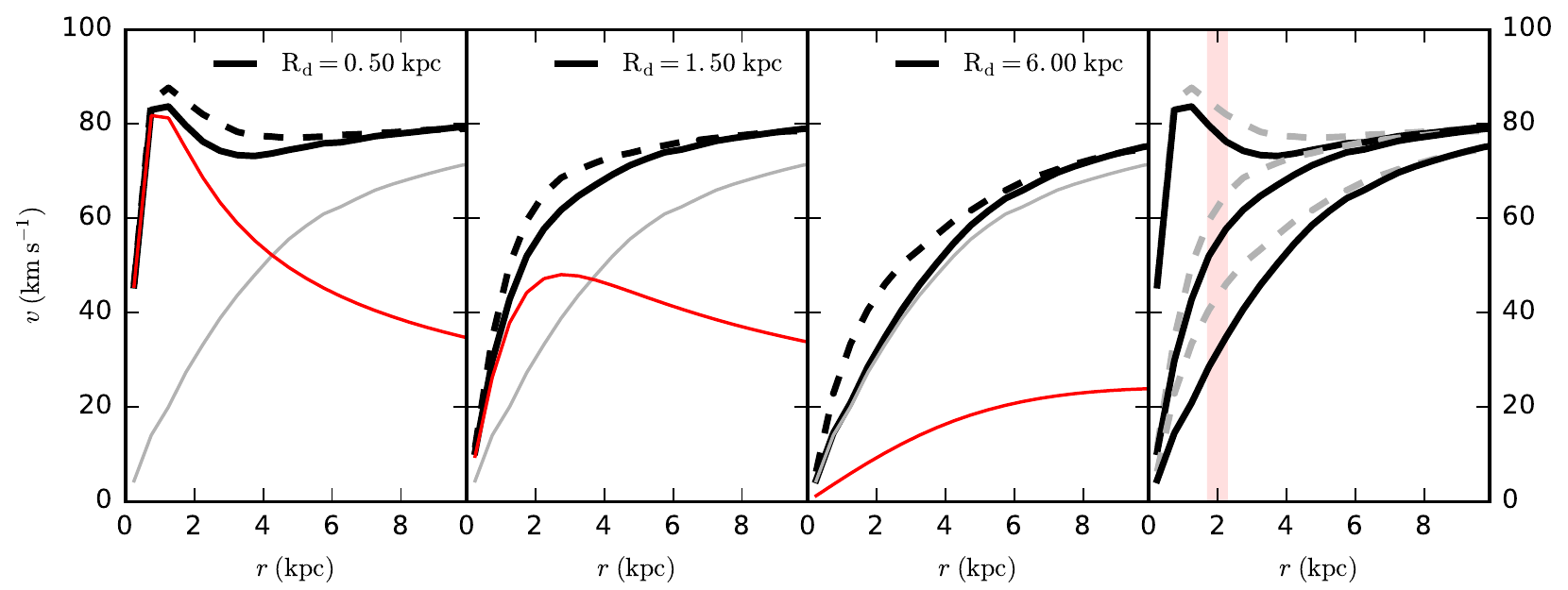}
  \caption{\emph{Left three panels:} Dark matter contribution to the circular velocity profile 
    for a dwarf galaxy ($V_{200}=70\; \kms$, $c=-2\sigma$)
    with a fixed baryonic disk component with scale radii
    $0.5$, $1.5$ and $6\; \rm kpc$ from left to right respectively,
    simulated for 10~Gyr with a cross-section $\som=2 \;\cmspg$.
    \emph{Red lines}
    indicate the baryonic contribution, \emph{grey} the
    dark matter (SIDM), and \emph{thick solid black} the total. For
    comparison, the \emph{thick dashed black} lines indicate the CDM
    total. \emph{Right panel} combines the three profiles, with the CDM
    comparisons in \emph{grey dashed lines}.}
\label{fig:vel_vs_rd}
\end{figure*}

\begin{figure*}
\includegraphics[width=\columnwidth]{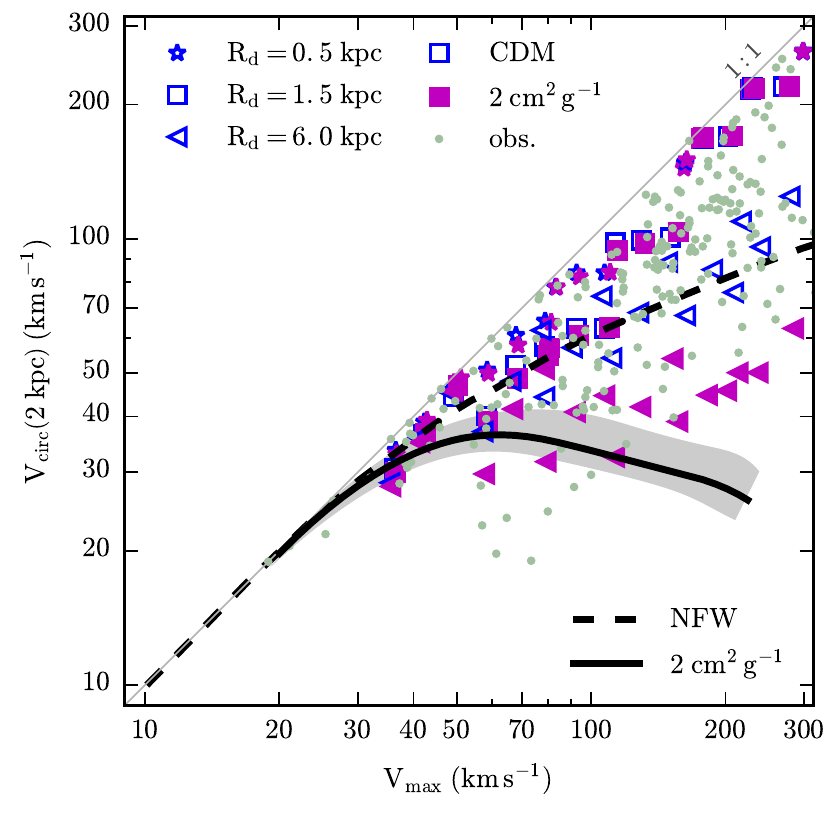}
\includegraphics[width=\columnwidth]{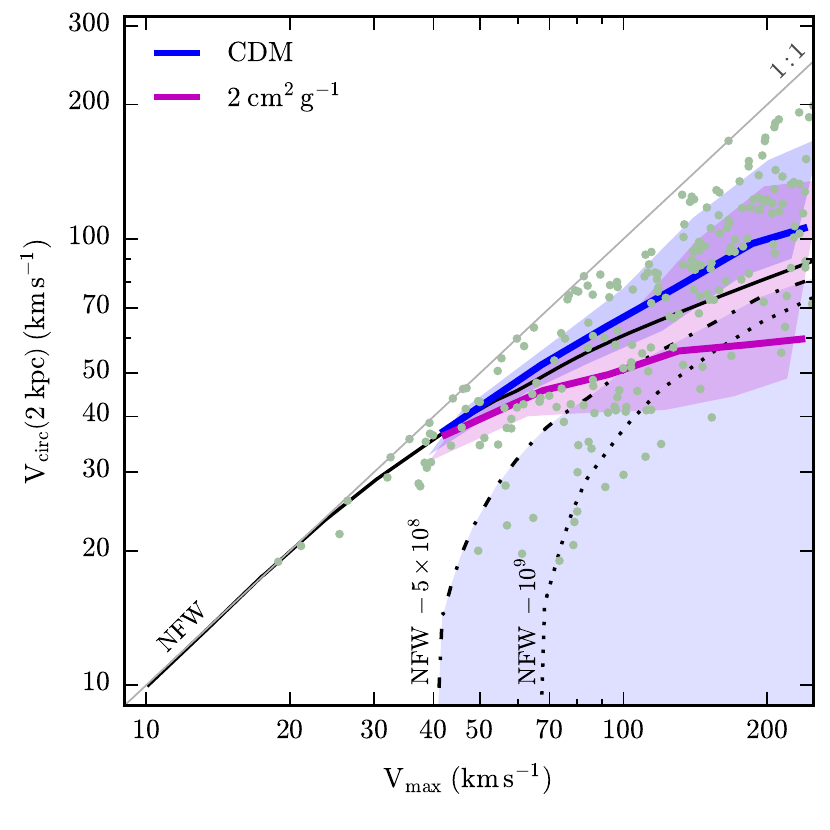}
\caption{N-body simulations in the presence of galactic disks. \emph{Left panel} shows
  the individual simulations with disk scale lengths $\rdisk=0.5$, $1.5$ and 6~kpc in \emph{stars}, \emph{squares} and \emph{triangles} respectively, where each is repeated for the three different
  $-2\sigma$,  median, and $+2\sigma$ concentrations and SIDM in \emph{solid magenta} and CDM in \emph{open blue symbols}. \emph{Dashed black line} indicates the CDM relation without baryons, and \emph{solid black} the corresponding SIDM (with $\pm 1\sigma$ shading). 
\emph{Right panel} shows 2-$\sigma$ likelihood contours about the median $\rdisk$ and concentration (\emph{solid lines}) for CDM in the presence of a baryonic disk in \emph{blue} and
  SIDM in \emph{magenta}, inferred from the fits in
  Eqns.~(\ref{eq:Shen_size})~and~(\ref{eq:Shen_size_scatter}), (see also main text). \emph{Green points} indicate observed values collated in Oman 2015 Table~1.}
\label{fig:SIDM_bar_distrib}
\end{figure*}

\subsection{Circular velocity profiles in the presence of baryons}\label{sec:V2_Vmax_bar}

For collisionless dark matter halos, the presence of baryons may
change the shape  of the {\it total} circular velocity profile by
adding the contribution from the gas and the stars. This contribution
may or may not change the overall profile, depending on the relative
fraction of baryonic mass and also the spatial distribution of these
baryons\footnote{Notice that baryons could
additionally cause the dark halo to contract \citep{Blumenthal_1986}, but given our
initial condition set-up, the halos will be created already in equilibrium
with the baryonic potential}. If gas and stars are very centrally
concentrated, they will dominate the contribution to the circular
velocity modifying the profile to rise more steeply than the original
NFW halo. On the other hand, if the baryons are fairly extended, their
contribution could be sub-dominant to the dark matter, in which case
the total velocity profile including baryons does not deviate
significantly from that of the original NFW halo. 

Fig.~\ref{fig:vel_vs_rd} (leftmost three panels) illustrates this
point using the N-body simulations set up described in
Sec.~\ref{sec:sims}. For a fixed halo with $V_{200}=70\; \kms$, 
$\mdisk = 2.8\times 10^9$~\Msun\ (according to abundance matching
relations from \citet{Behroozi_2013}) and a concentration $-2\sigma$ below the
median.
We plot the equivalent  `circular velocity' inferred for a spherically
averaged mass distribution
\begin{equation}\label{eq:vcirc}
V_{\rm cir}^2 = G M(r)/r = V_{\rm cir, dm}^2 + V_{\rm cir, bar}^2
\end{equation}
which is an approximation for non-spherical distributions
(for a pure thin MN disk without a halo these 
 deviations are at maximum $\approx 15\%$).
We assume that
the baryons distribute in a MN disk with scale length
equivalent $\rdisk = 0.5$, $1.5$ and $6\; \rm kpc$ (left to right). The
long-dashed lines indicate the final profiles including the baryons
for CDM, which show different shapes according to $\rdisk$, with the
differences tracing back to the contribution of the baryons in each
case.

As expected, a more concentrated baryonic distribution (small $\rdisk$)
shows a steep raise of the circular velocity curve, whereas larger
$\rdisk$ values grow more gently. The case with $\rdisk=0.5$~kpc results in
a declining circular velocity profile which might not be the typical
galaxy included in studies of rotation curves (probably representing a
bulge dominated object instead). However, we include this case for
illustration of the possible extremes and how CDM and SIDM would react to
the presence of such a compact galaxy.

The same exercise of considering a fixed halo and baryonic content
but changing the spatial distribution can be done assuming the SIDM
scenario. The three left panels in Fig.~\ref{fig:vel_vs_rd} show the
expected total profiles assuming a self-interaction cross-section
$\som=2 \; \cmspg$ (see solid black lines). The dark matter component still
shows a cored profile due to the self-interactions (see thin grey
line) but this is obscured in the \emph{total} velocity profile. In fact,
the inclusion of baryons creates a fair amount of {\it diversity} in
the shapes of the circular velocity profiles for CDM and SIDM (see
right panel in the same figure). For example, if we measure $\Vtwo$ in
these halos (vertical red shaded region), CDM covers a range $\approx
45-83$ $\kms$ which is comparable to that spanned by SIDM, $\Vtwo
\approx 30-75$ $\kms$. Encouragingly, the effect of baryons in the case
with self-interactions creates a wide range of (total) rotation curves but
also maintains a lower central density (in the cases where baryons are
extended), allowing better accommodation of the low $\Vtwo$ values
that are occur in observations (see right panel
Fig.~\ref{fig:dm_only}) and for which CDM alone has no explanation.

We generalise this by simulating all halos introduced in
Sec.~\ref{sec:V2_Vmax_dm} but including a baryonic disk. The mass of the
disk is the sum of a stellar and gaseous component, with the first set
by the abundance matching relation between $M_*$ and $M_{200}$
introduced in \citet{Behroozi_2013} and the gas mass computed from the
$M_{\rm gas}$-$M_\star$ relations of \citet{Huang_2012}. To account
for variations in halo concentration $c$, we sample three different halos
of median and $\pm 2\sigma$ extremes according
to the mass-concentration relation from
\citet{Ludlow_2014}. Table~\ref{table:bar_param} summarises the main
properties of all our runs. Each halo is then set up in equilibrium
with their baryonic MN disk of scale lengths
$\rdisk=0.5,\; 1.5,\; \rm and \; 6.0$~kpc. We fix the scale
height of the disks to $\zdisk = 0.3\rdisk$, a slightly hotter
structure than for Milky-Way-like galaxies but that provides a reasonable
description of lower mass galaxies, the main target of this study.

We run our simulations for $10$ Gyr assuming $i)$ collisionless cold dark
matter (CDM) and $ii)$ a fiducial self-interaction cross-section
$\som = 2\;\cmspg$ for the SIDM case.  We investigate the relation
between $V_{\rm max}$ and $\Vtwo$ of our halos in CDM and SIDM in the
left panel of Fig.~\ref{fig:SIDM_bar_distrib}. Symbol shapes (square,
star, triangle) indicate the different scale lengths sampled whereas
open-blue and filled-magenta differentiate between CDM and SIDM
runs. A quick inspection suggests that self-interactions are able to
generate a wider range of $\Vtwo$ at fixed $V_{\rm max}$ which is in
better agreement with observed values from the compilation in
\citet{Oman_2015} (green dots).  The diversity in both cases, CDM and
SIDM, arises from different contribution by the baryons.  Since SIDM
cores imply a lower dark matter density in the inner regions, the
contribution of baryons is \emph{more} important that in the
collisionless case, creating a larger variety in $\Vtwo$. Notice that
for the more compact disks ($\rdisk=0.5$ and $1.5$ kpc) the
predictions from CDM and SIDM are not that different, and in both
cases $\Vtwo$ can be very close to $\Vmax$, which suggests these
rotation curves are relatively flat. One should also note that
in a cosmological context the central density would be even further
promoted by the adiabatic contraction introduced by baryonic collapse
\citep[e.g.][]{Elbert_2016}, which occurs even in the models with
bursty feedback \citep{Tollet2016}.
In the case with non-dominant
baryonic components (triangles) the points stay close to the dark
matter-only case (thin solid lines), which is significantly lower in
the case of self-interactions, as discussed in
Sec.~\ref{sec:V2_Vmax_dm}. 

The lower envelope of points in either scenario corresponds to the
most extended disks populating the $-2\sigma$ outliers in halo
concentration. A few of the observed points in the left panel of
Fig.~\ref{fig:SIDM_bar_distrib} still remain unexplained
($\Vmax \approx 70 \; \kms$ and $\Vtwo \approx 20\; \kms$) showing
central densities even lower than can be obtained in SIDM with
$\som=2\;\cmspg$. This could be suggesting the need for a larger cross
section for self-interactions or more extreme outliers on the
mass-concentration relation. We should also bear in mind the
possibility of observational errors as discussed in Sec.~\ref{sec:intro}.

Although the exercise of fixing $\rdisk$ is more clear and intuitive,
in practise we know that galaxy size will scale with stellar mass, and
the average behaviour in $\Vmax$-$\Vtwo$ plane will depend on the
average $\rdisk$ of the population at fixed $\Vmax$.  In the right
panel of Fig.~\ref{fig:SIDM_bar_distrib} we take this into account as
follows.  The first step is to characterise $\rdisk$ as a function of
$\Vmax$. Because of the tight relation between halo mass and stellar
mass, this is equivalent to finding a relation between $\rdisk$ and
$M_*$ or, in our case, $\mdisk$. We compute the stellar size of the
disk and then estimate the baryonic one by a simple correction by gas
fraction. In detail, for our stellar radii we use a fit to the
late-type SDSS galaxies of \citet{Shen_2003}:

\begin{equation}\label{eq:Shen_size}
  R_{\star} = 0.1  \, \left( \frac{M_\star}{1 \; M_\odot } \right)^{0.14} \left( 1 + \frac{M_\star}{3.98 \times 10^{10} \; M_\odot} \right)^{0.25} \; \rm kpc
\end{equation}

\noindent
with scatter
\begin{equation}\label{eq:Shen_size_scatter}
  \sigma_{{\rm ln} R} = 0.34 + \frac{0.13}{1 + \left( \frac{M_\star}{3.98 \times 10^{10} \; M_\odot} \right)^2}
\end{equation}

\noindent
along with an estimate of the baryonic size as: $\rdisk = (1+f_{\rm
  gas}) R_\star$, where $f_{\rm bar}$ is estimated from $M_*$ and
observations by \citet{Huang_2012} as explained before. 
This procedure then provides a median and $\pm
2\sigma$ disk radii for each halo mass. We then take these dispersions
in $\rdisk$ and interpolate between our simulations to find the
appropriate $\Vmax$ and $\Vtwo$, finally combining this dispersion in
quadrature with the dispersion due to the $M_{200}$-concentration
relation to infer the median and $\pm2 \sigma$ likelihood regions as a
function of halo mass.

These likelihood regions are shown in the right panel of
Fig.~\ref{fig:SIDM_bar_distrib}. The expected median in each scenario
is notably shallower than the points in the left panel, which is
mainly due to the average disk size rising as $\Vmax$
increases. Indeed, as we saw in the left panel of the same figure, at
fixed $\Vmax$ a more extended baryonic disk will have lower
$\Vtwo$. For comparison we include the effects of the mass deficits of 
$5\times 10^8$ and $10^9$\Msun within 2~kpc.
Several observed points still scatter above and below the median
relations for both CDM and SIDM hinting at the need of more extreme outliers
(more than $2 \sigma$) to explain such objects. 

We have probably overstated the extremity of these outliers since we 
have at zero-th order ignored correlations between halo and disk properties 
that would alleviate the tension. 
For instance, if more compact disks systematically
populate more concentrated halos whereas extended disks live in low
concentration halos, the shaded areas could be larger. We have also not 
made an attempt to account for observational biases in these
calculations. Regardless of the absolute value, however, we emphasise that at
fixed assumptions, SIDM seems to always provide a larger
range of diversity than CDM (with scatter at least $150\%$ of the latter).

\begin{figure}
\includegraphics[width=\columnwidth]{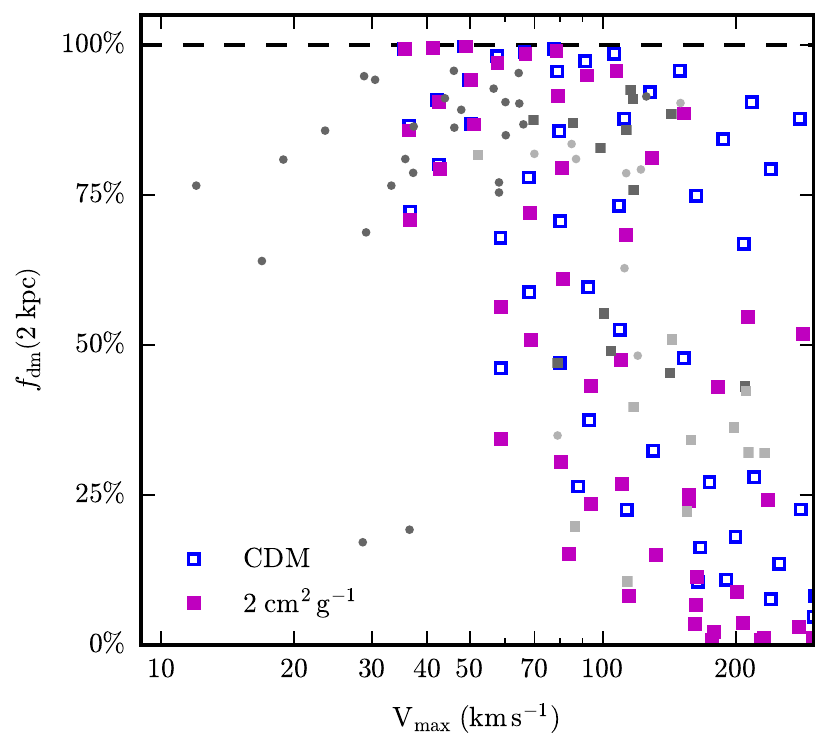}
\caption{Dark matter contribution to the rotation curve at 2~kpc for SIDM
  simulations and observed dwarfs (see main text). \emph{Filled magenta squares}
  are the SIDM simulations, in comparison to the CDM in \emph{open blue squares}.
  \emph{Greyscale symbols} refer to observed
  galaxies from \citep{Oh_2015} (\emph{dark circles}),
  \citet{Kuzio_2008} (\emph{light circles}), \citet{deBlok_2001} 
  (\emph{light squares}) and \citet{deBlok_2008} (\emph{dark squares}).}
\label{fig:DM_2kpc}
\end{figure}

An interesting corollary of the analysis above is that, fixing the
relation $\Vmax$-$\Vtwo$, a given halo in CDM and SIDM would be
predicted to have different contribution
from the baryons at $2$~kpc, since in SIDM the baryonic matter will have to
``compensate'' for the lower density of the cored dark matter halo. It
is therefore important to compare the dark matter fractions predicted
in both models to check that they are consistent with observed
galaxies. Fig.~\ref{fig:DM_2kpc} shows the dark matter
fractions $f_{\rm dm}=M_{\rm dm}/M_{\rm tot}$ measured at $2$~kpc for
our CDM and SIDM halos (open and filled symbols, respectively)
compared to a subsample of galaxies taken from the rotation curves
presented in \citep{Oh_2015} and \citet{Kuzio_2008}.  For the former
we use their disk-halo decomposition from the stellar, HI and
kinematic data, and for \citet{Kuzio_2008} we use their `popsynth'
mass modelling rather than the extremal fits. In a couple of cases the
last-measured point does not reach 2~kpc and for these we used the
dark matter fraction at the last measured point. 

There is a significant overlap between the collisionless and the 
self-interacting scenarios. Also, spatial resolution coupled to
uncertainties in the mass-to-light ratios for the observed objects
makes $f_{\rm dm}$ probably not a good enough indicator to distinguish
between these alternative models. Nonetheless, it is reassuring that
the same SIDM objects that reproduce better the observations in the
$\Vmax$-$\Vtwo$ plane, seem to have dark matter fractions that are
consistent with current observations.

\subsection{Two extreme examples: IC~2574 and UGC~5721}\label{sec:obs_curve_sims}

In Sec.~\ref{sec:V2_Vmax_bar}, we presented a detailed statistical
analysis of a sample of objects simulated within the SIDM
paradigm. This, however, left unexplained one of the main motivations
of this work, which is the existence of extreme outliers which are not
possible to accommodate within the standard CDM scenario. We turn then
our attention to two particular examples highlighted in
\citet{Oman_2015}, UGC~5721 and  IC~2574. These two galaxies have
similar circular velocity measured at the outermost point of the
rotation curve $\approx 80$ $\kms$, which probably indicate that they
populate dark matter halos with similar mass, but the {\it shapes} of
the inner regions are remarkably different: whereas UGC~5721 is
consistent with a `cuspy' NFW profile, IC~2574 has a very extended core.
This pair is an interesting case
of the aforementioned diversity and we can use it as benchmark for our
SIDM scenario.

\begin{figure}
\includegraphics[width=\columnwidth]{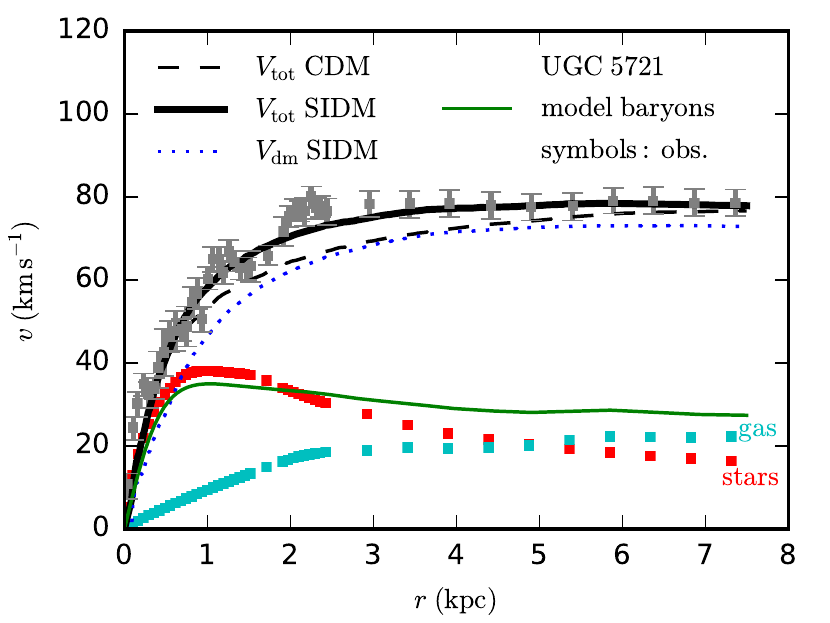}
\caption{Observations for UGC~5721 along with SIDM and CDM simulations. \emph{Red points} indicate the stellar contribution, \emph{cyan} the gas and \emph{grey squares with error bars} the observed rotation curve. The SIDM simulation (with parameters from Table~\ref{table:gal_params_fits}) has the baryonic mass contribution indicated by the \emph{green line}, which is a composition of the interpolated gas distribution along with an MN fit to the stellar disk. The halo contribution is given in the \emph{dotted blue line} and the total curve in \emph{solid black}. The comparison curve for CDM is given in the \emph{dashed black line}.}
\label{fig:UGC5721}
\end{figure}

\begin{figure}
\includegraphics[width=\columnwidth]{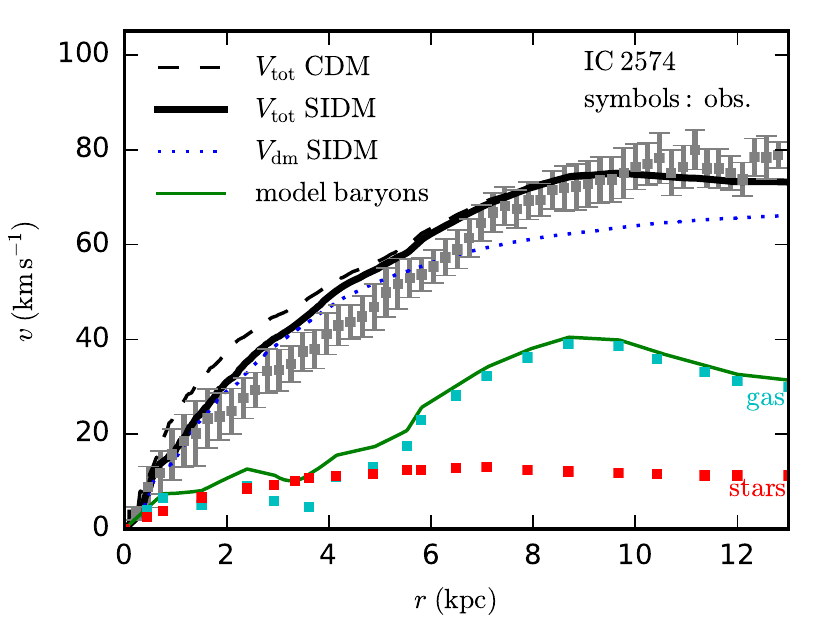}
\caption{Observations for IC~2574 along with SIDM and CDM simulations. \emph{Red points} indicate the stellar contribution, \emph{cyan} the gas and \emph{grey squares with error bars} the observed rotation curve. The SIDM simulation (with parameters from Table~\ref{table:gal_params_fits}) has the baryonic mass contribution indicated by the \emph{green line}, which is a composition of the interpolated gas distribution along with an MN fit to the stellar disk. The halo contribution is given in the \emph{dotted blue line} and the total curve in \emph{solid black}. The comparison curve for CDM is given in the \emph{dashed black line.}}
\label{fig:ICgal}
\end{figure}

In Figs.~\ref{fig:UGC5721} and \ref{fig:ICgal} we show the data together
with the best fit CDM (dashed line) and SIDM (solid black) curves. 
For UGC~5721 (NGC3724) we use the rotation curve data
from \citet{Swaters_2003} and stellar and gas densities from Andrew
Pace (priv. communication, for comparison one can see
Fig.~1 in \citealp{Swaters_2010} which has a slightly higher
mass-to-light ratio).

For IC~2574 we use the rotation curve, gas and stellar densities from
\citet{Oh_2011}. We multiply the stellar velocity contribution by
$0.5$ (mass-to-light ratio lower by a factor $4$), so the stellar
contribution lies between that in \citet{Sanders_2002} and
\citet{Oh_2011}. For both cases our SIDM halo is run using a cross
section for self-interactions of $\som = 3\,\cmspg$
following \citet{Kamada2016}.

\begin{table}
\begin{center}
\begin{tabular}{lccc}
\hline

Property & Symbol & IC~2574 & UGC~5721 \\
\hline
Disk scale length & $\rdisk$ & $3.0$ & $0.50$ kpc \\
Disk scale height & $\zdisk$ & $0.9$ & $0.15$ kpc \\
Stellar mass & $M_\star$  & $3.00 \times 10^8$ & $4.48\times 10^8$ \Msun\\
Gas mass$^\dagger$ & $M_{\rm gas}$ & $2.72\times10^9$  & $8.47\times10^8 $ \Msun\\
Halo mass & $M_{200}$ & $9.78\times10^{10}$ & $5.15\times 10^{10}$ \Msun \\
Concentration & $c$ & $6.2$ ($-2.3\sigma$) & $13.6$ ($+0.9\sigma$) \\
DM cross-section &$ \som $ & $3$ & $3$ cm$^2$ g$^{-1}$ \\
NFW radius & $r_{\rm s}$ & $15.4$ & $5.65\;\rm kpc$ \\
NFW density & $\rho_{\rm s}$ & $1.9\times10^{-3}$ & $0.013 \, \rm M_\odot\, pc^{-3}$ \\
\hline
\end{tabular}
\end{center}
\caption{Parameters for the SIDM halos and baryonic potentials for the two galaxies in Fig.~\ref{fig:UGC5721} and \ref{fig:ICgal}. \newline 
$\dagger$ The gaseous component as added in post-processing due to its disturbed nature, see also main text in Sec.~\ref{sec:obs_curve_sims}. \newline 
}
\label{table:gal_params_fits}
\end{table}

Our fitting process relied upon making an initial guess for the 
halo mass using $\Vmax$ and sampling the concentration with $\pm 2.5 \sigma$ of the 
mass-concentration relation in Eqn.~(\ref{eq:linear_ludlow}). 
The combined stellar plus gaseous disks in these dwarfs are not
well represented by a single MN disk (especially their gas profiles
which exhibit a much more extended, nearly constant surface density), 
however since these differences are primarily in the outer regions where we
expect the effects of self-interactions to be sub-dominant, we chose
to approximate only the \emph{stellar} distribution with an MN disk 
and apply the gas contribution to the rotation curve in a 
post-processing step (i.e. add in quadrature as in Eqn.~\ref{eq:vcirc}),
and after several iterations with the disk component we find the 
parameters indicated in Table~\ref{table:gal_params_fits}.
As noted in Appendix~\ref{sec:NFW_vs_HQ},
the initial conditions are constructed with Hernquist profiles 
converted from the corresponding NFW parameters, 
but if one is using a real NFW then in the higher
concentration case using a lower $r_{\rm s}$ would give a
slightly better match.

Fig.~\ref{fig:UGC5721} shows that the SIDM fit for UGC~5721 including all components has an
overall ``cuspy'' profile (see black solid line), which is \emph{steeper} than the circular
velocity expected for the corresponding NFW halo without 
self-interactions (dashed line). For this fit we need a concentrated
halo at $+1\sigma$ above the median mass-concentration relation, which
together with the observed compact stellar disk (scale radius of
$0.5$~kpc) combine to create a steeply raising velocity curve. 
This is an excellent example of the importance of baryons in the shape
of the rotation curves, even for dwarf galaxies. 
The $\chi^2/\rm dof$ of the SIDM fit is $3.04$ ($4.44$ below the initial, 
which is of course not the best individual CDM halo fit). 
The visual impression of a superior fit is likely due to 
This $\chi^2$ is largely driven by the $\lesssim 0.3$~kpc data with
tight error bars over a non-monotonic feature, contributing to 
a visually impression of a superior fit.
Clearly a more accurate fit to such features would require additional parameters (particularly for the stellar components), however this is outside the scope of this paper.
Without the presence of baryons, UGC~5721 is a challenge for SIDM
due to the development of an extended low density core at the centre. In our simulations, however, the
SIDM thermal transfer in cooperation with the baryon dominance in the
central $1.5$~kpc causes the halo to shrink even further than the
pure NFW model. 
The compact stellar component excavates a potential well sufficient that
thermalisation of the dark component leads to an almost-NFW final
configuration \citep[see also ][]{Kamada2016}.

On the other hand, Fig.~\ref{fig:ICgal} presents the rotation curve
fit for IC~2574 with an extended and slowly rising core in its
centre. The gas has a disturbed profile, but the baryons as a whole
contribute little to the total density except within the innermost
$0.5$~kpc, and make a modest contribution at $r> 6$~kpc. For such a
dark matter dominated galaxy to deviate so far from a cusp-like
profile is clearly a challenge for vanilla CDM \citep{BlaisOuellette_2001, Sanders_2002,
  Swaters_2003, Swaters_2010}. Notice that any baryonic solution to
this problem within CDM -- for instance assuming that supernova
feedback can create a core-- would not work for such object, since the
core extends well beyond where the stars are (see discussion in
\citealt{Oman_2015}). 
The $\chi^2/\rm dof$ of the SIDM fit is $1.57$, ($1.39$ below the initial
conditions), which reflects the introduction of the core
and quantifies the capacity of SIDM as a solution to this conundrum.

Whilst SIDM is relatively efficient in producing cores, this process
becomes increasingly sedate as the core size approaches the scale
radius of the halo (in this case $15.4$~kpc) and so we have had to
choose a relatively large scale radius (i.e. low concentration)
halo. The $\Vtwo$ of our CDM halo is around $35\; \kms$ whilst in SIDM
it is around $27\; \kms$, bringing it in better agreement with the
observed rotation curve $\Vtwo \approx 23 \; \kms$. 

We note that for IC~2574 both the low concentration and self-interactions are required to achieve a good fit. As such one way to test our model is to perform a statistical study of extreme outliers like IC~2574. Since we assume the SIDM model inherits the halo mass-concentration relation of the CDM one, such a study also provides a direct test for \lcdm\ on galactic scales.

We have shown that the SIDM model in the presence of baryons can provide good fits to two extreme outliers. We demonstrate that with SIDM the halo concentration, the baryon concentration and the back-reaction of baryons on the density profile play roles in explaining the rotation curves of these two extremes. SIDM halo profiles are flexible enough to accommodate the diverse rotation curves of spiral galaxies.

\section{Summary and Conclusions}\label{sec:discussion}

The SIDM model modifies the CDM paradigm by inclusion of non-gravitational interactions between dark matter particles in the halo evolution, and these self-interaction effects are most significant in the regions of halos with the high densities and velocity dispersions, i.e. within the scale radius of the halo. In the outer halo and at large scales SIDM is almost collisionless and retains the successes of \lcdm.
In the inner regions of halos the dark matter is redistributed into a central core, whose density and size depend on the microscopic particle interaction cross-section, the halo mass and even the baryon concentration. The mass profiles inferred from astronomical observations of the rotation curves of spiral galaxies thus offer an avenue to both constrain the self-interaction cross-section of dark matter and falsify or support specific self-interaction models.

Observationally, the shapes of the rotation curves of galaxies exhibit a
wide diversity. We follow \citet{Oman_2015} by quantifying the range in the rotation curves using the relation
between the total circular velocity measured within $2~{\rm kpc}$ and at the peak, $\Vtwo\textup{--}\Vmax$. For a given $\Vmax$, the spread in $\Vtwo$ can be a factor of four, which is difficult to reconcile in the prevailing \lcdm~paradigm~\citep{Oman_2015}, where dark matter is assumed to be collisionless over the cosmological timescale. This problem stems from the hierarchical structure formation in \lcdm\ which produces a self-similar halo density profile, which is essentially parameterised by one parameter, the halo mass (or concentration). After we determine the halo mass by $\Vmax$, the halo circular velocity is completely fixed at all radii, up to the scatter, including the inner density cusp that is in contrast to the shallow (cored) one preferred by many dwarf galaxies. In addition, the enclosed mass of a CDM halo cusp tends to overwhelm the baryonic contribution and consequently also the scatter in $\Vtwo$ caused by the spread in the baryon concentration.
This inflexibility in CDM may be alleviated by introducing particular prescriptions of baryonic feedback processes that can dynamically heat the dark matter \citep[and others]{Navarro_1996,Pontzen_2012}, however SIDM offers an attractive alternative solution.

In this paper we have used controlled N-body simulations as a numerical experiment to test the SIDM model's ability to address the diversity problem. We use the {\sc Arepo} code to run isolated simulations of a live halo in the presence of a static baryonic disk in order to assess the gravitational impact of gas and stars on the halo evolution. This approach is suited to the SIDM model as dark matter self-interactions thermalise the inner halo and the final (inner) density profile is largely insensitive to the formation history. We sample a wide range of halo masses spanning about three orders of magnitude and vary the halo concentration within $\pm2\sigma$ from the its mean value. For each halo mass, we use abundance matching relations to choose the disk mass and we take three disk scale radii to span the baryon distributions. This sampling of the model parameter domain enables us to make concrete predictions of the rotation curves in the SIDM model. 

Our main conclusion (see Fig.~\ref{fig:SIDM_bar_distrib}) is that the SIDM mechanism accommodates galaxies with a wide range of  $\Vtwo$ over the domain of $\Vmax$ from $30$--$250\;\kms$. The spread in core densities is due in part to the variance in concentration (within $\pm2\sigma$ from the mean), however this alone is insufficient to account for the large relative scatter of the observed values. Including self-interactions allows lower $\Vtwo$, which requires low central densities in \emph{both} baryons and dark matter. High values of $\Vtwo$ are still achieved with compact baryonic disks where thermalisation into the deep baryonic potentials can induce NFW-like densities. 
As a result the predicted range of $\Vtwo$ for a fixed $\Vmax$ in SIDM is $50\%$ larger than the CDM one, leading to a better agreement with the observed rotation curve distribution.

Some observed galaxies in the range $\Vmax=60$--$80\; \kms$ and $\Vtwo=20$--$30\;\kms$ are still challenging.
These cases might require 
assumptions about the structure of the copula of the combination of 
halo concentrations and baryonic disk sizes, in particular that very
extended disks populate the most under-concentrated halos to reproduce the very low $\Vtwo$, which is not an 
unrealistic assumption to first order. 
Nevertheless, the SIDM prediction is an improvement over CDM in all cases.

To further test our results, we have performed simulations to provide individual fits for two of the most extreme
outliers highlighted in~\citet{Oman_2015}, UGC~5721 and IC~2574, cusp- and core-
dominated galaxies respectively, both with similar $\Vmax \approx
80 \; \rm \kms$ (see Figs.~\ref{fig:UGC5721} and~~\ref{fig:ICgal}). Picking extreme $+0.9\sigma$ and $-2.3
\sigma$ concentrations from the halo concentration-mass
relation, together with a compact $\rdisk = 0.5~{\rm kpc}$ and
extended $\rdisk = 3.0~{\rm kpc}$ baryonic distribution respectively, we find good fits in the SIDM paradigm. Interestingly, we find that the SIDM density profile can be as dense as the NFW one in a baryon-concentrated galaxy such as UGC~5721, because SIDM particles follow an isothermal distribution within the deep baryon potential. Our simulation result confirms the theoretical expectation first predicted in~\cite{Kaplinghat_2014}. 
At the other extreme, IC~2574 with an inferred central dark matter dominated core is more straightforwardly tractable with a model that allows dark matter thermalisation, and is more challenging to the collisionless CDM model.

The implication of these elements is that with the inclusion of realistic baryonic components, SIDM models can produce quantitatively superior fits to rotation curves, both at the level of individual rotation curves and over the statistical distribution of a quantitative measure of their shapes. As one would expect, fitting the most extreme outliers requires an interplay between the combinations of halo concentration, the baryon distribution, and the influence of baryons on the SIDM halo profiles. The results from our numerical experiments fit well with the previous theoretical calculations of \citet{Kamada2016}. 

A number of avenues for future research are apparent. On the observational side, more detailed observations for mass modelling would allow tighter constraints on the dark matter distribution and consequently the interaction cross-section. 
Orthogonally, larger sample sizes would allow the $\Vtwo$--$\Vmax$ distribution to better discriminate between CDM and SIDM. 
On the computational side, we have ignored the effects of the hierarchical assembly of dark matter halos. Whilst this is not expected to play a large role in the inner halo, it will nevertheless introduce additional scatter as a function of assembly time. Finally, all these will likely show strong correlations with the baryonic statistics, and investigations of galaxy formation in SIDM using full hydrodynamical cosmological simulations with the baryonic feedback processes are already being explored \citep[e.g.][]{Vogelsberger_2014,Fry_2015}.

\section*{Acknowledgements}
The authors would like to thank Andrew Pace for sharing his collated data set for the rotation curves, 
Volker Springel for making {\sc Arepo} available for this work and
James Bullock, Simon White, Oliver Elbert, and Manoj Kaplinghat for useful discussions. This work was supported by the U.S. Department of Energy under Grant  No. de-sc0008541 (HBY). HBY acknowledges support from the Hellman Fellows Fund. MV acknowledges support through an MIT RSC award and the support of the Alfred P. Sloan Foundation.

\bibliographystyle{mnras}

\bibliography{paper}

\appendix
\section{Jeans analysis}\label{sec:jeans}

In order to accelerate the analysis of the spherically symmetric cases we have employed `Jeans analysis' \citep{Kaplinghat_2016} which gives a very good approximation to the final density distribution of halo-like systems. For the benefit of the astrophysics reader we discuss the self-similar evolution of the dark matter only case here, for more complex cases one should consult the aforementioned reference.

For the spherically-symmetric dark matter only case, the distribution function of dark matter can be written as $f(r, v_\parallel, v_\perp)$ with density $\rho_{\rm ini}(r)$. An approximation to the effects of self-interaction, referred to as Jeans analysis, is to assume that those regions of the halo with $\lesssim 1$ scattering over the evolution are unperturbed, whilst those with $\gtrsim 1$ have become fully collisional (i.e. act as an ideal gas) and isothermal (i.e. erased any thermal gradient). Whilst this approximation may seem relatively severe, one should recall that a dark matter halo encompasses many orders of magnitude in density and velocity dispersion, and so the majority of the volume is either highly collisional or nearly collisionless. Finding the radius $r_1$ at which approximately 1 scattering per evolution time $T$ occurs is determined implicitly via
\begin{equation}\label{eq:r1}
\frac{\left<\sigma_{\rm T} v\right>}{m_\chi} \rho_{\rm ini}\left(r_1 \right) T = 1\, ,
\end{equation}
and subsequently the evolved profile (assuming $f_{\rm ini}$ was a stationary state) is the piecewise solution
\begin{equation}
\rho(r)=\left\{ \begin{array}{cc}
\rho_{\rm iso} \left( r \right) & r<r_1 \\
\rho_{\rm ini} \left( r \right)  & r\geq r_1 \end{array} \right.
\end{equation}
where $\rho_{\rm iso}(r)$ is the isothermal solution to the Jeans equation (see below) with $\rho_{\rm iso} \to$ a constant as $r\to 0$ that matches the density and mass enclosed $\rho_1$ and $M_1$ at $r_1$. 

\subsubsection{Solution for $\rho_{\rm iso}$}

Consider Poisson's equation for an isothermal self gravitating spherically symmetric halo
\begin{equation}\label{eq:pois}
\sigma_0^2 \nabla^2 \ln \rho = -4\pi G \rho
\end{equation}
As a second order differential equation this has 2 unknowns, which extends to a third if we consider $\sigma_0$ (the velocity dispersion) to be unknown. If we know two values, e.g. $\rho_1$ and $M_1$ at some radius $r_1$, then we need an additional constraint, but this can be provided by choosing the `cored' solution with $\rho \to$ a constant as $r \to 0$. 

Denoting this constant $\rho_{\rm c}$ a good choice for similarity function turns out to be for the ratio of the mean enclosed density $\bar{\rho}$ to this, i.e. $\theta = \frac{\bar{\rho}}{\rho_{\rm c}}$.  Expanding the Taylor series about $r=0$ tells us the leading order non-constant term is $O(r^2)$, i.e. w.l.o.g we can write $\theta \sim 1-\frac{r^2}{r_{\rm c}^2}$ (for some core size $r_{\rm c}$) as $r \to 0$. This suggests the use of the dimensionless variable
\begin{equation}
\mu = \frac{r^2}{r_{\rm c}^2} \, ,
\end{equation}
as a parameter for $\theta(\mu)$, and by substitution
\begin{equation}
\theta(0) = -\theta'(0) = 1
\end{equation}
where the prime denotes differentiation w.r.t. $\mu$, and we have chosen $r_{\rm c}$ to satisfy
\begin{equation}
\sigma_0^2 = \frac{2\pi}{5} G r_{\rm c}^2 \rho_{\rm c} \, .
\end{equation}
By substitution into Eqn.~(\ref{eq:pois}) we determine that $\theta(\mu)$ obeys the 2nd order nonlinear ODE
\begin{equation}
\theta = -\frac{3}{5} \frac{\rm d}{{\rm d}\mu} \ln \left(  3\theta + 2\mu \theta'\right) \, .
\end{equation}

\begin{figure}
\includegraphics[width=\columnwidth]{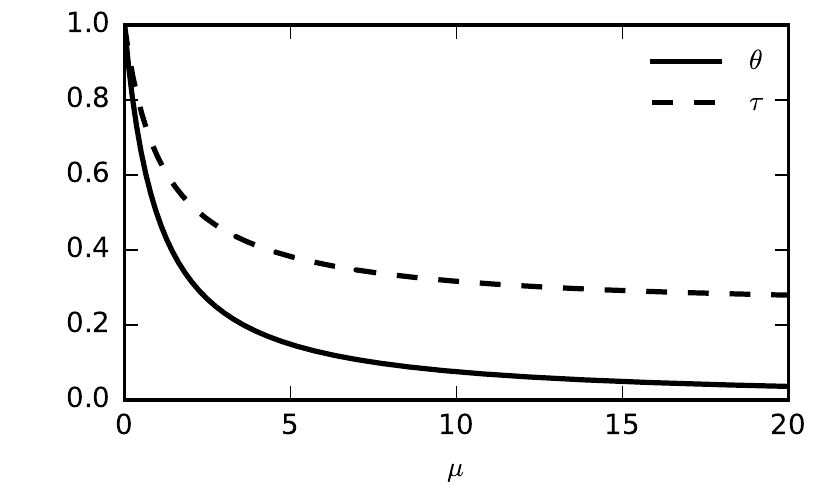}
\includegraphics[width=\columnwidth]{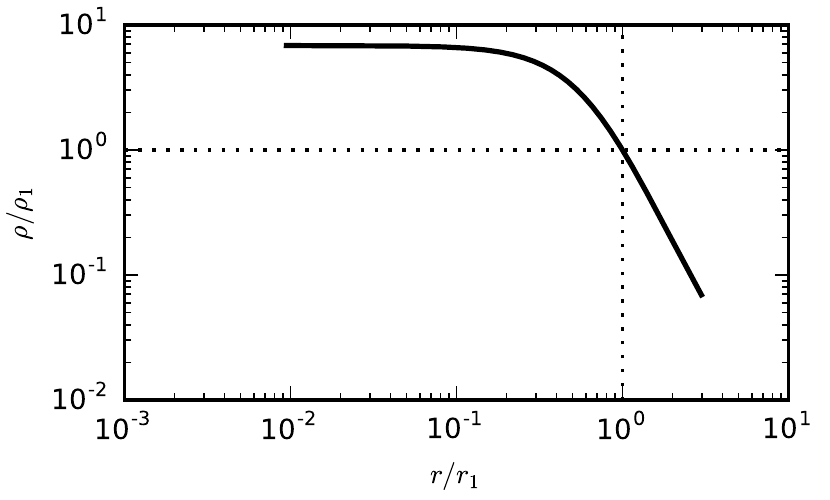}
\caption{\emph{Upper panel:} $\theta(\mu)$ and $\tau(\mu)$ functions. \emph{Lower panel}, example Jeans solution for $\tau_1=\frac{1}{2}$, i.e. at $r_1$ the density is half the mean density}
\label{fig:theta}
\end{figure}
We plot this function in the left panel of Fig.~\ref{fig:theta}. In order to determine the density we may wish to tabulate $\theta$ along with the dimensionless density ratio function we will refer to as $\tau$, determined from $\theta$ by
\begin{equation}
\tau(\mu) = \frac{\rho}{\bar{\rho}} = 1+\frac{2\mu \theta'}{3\theta} \, .
\end{equation}
which is monotonically decreasing. 

Given some point $r_1$ at which we know the density $\rho_1$ and the mass enclosed $M_1$, then we know the mean density and can simply read off the $\mu_1$ that gives us this $\tau_1$, i.e.
\begin{equation}
\tau(\mu_1) = \tau_1 = \frac{4\pi r_1^3 \rho_1}{3 M_1}
\end{equation}
and then substitute to find the density at any $r<r_1$, i.e.
\begin{equation}
\rho_{\rm iso}(r) = \bar{\rho} \tau = \rho_1 \frac{\tau\left( \frac{\mu_1 r^2}{r_1^2} \right)}{\tau_1} \frac{\theta\left(\frac{\mu_1 r^2}{r_1^2} \right)}{\theta_1} \, .
\end{equation}
and we plot an example in the right panel of Fig.~\ref{fig:theta}

\subsubsection{Scale dependence}
Since the halos in cosmological simulations are self-similar, the features are broadly speaking common to all halo masses. In particular in Eqn.~(\ref{eq:r1}), for a fixed halo concentration (at all halo masses) the density $\rho_{\rm ini}\left(r/r_{200} \right)$ is fixed, and so the scale dependent differences occur due to the increasing velocity dispersions and falling concentrations of larger halos. Increasing velocity dispersion increases the scattering rate and causes core formation to progress faster, whilst lower concentrations reduce the central densities, and consequently the scattering rate and core formation. Overall the velocity dispersion effect dominates and larger halos tend to grow their cores faster, up to some limit where velocity dependent cross-sections become important, although we do not discuss that case here.

\section{NFW vs. Hernquist ICs}\label{sec:NFW_vs_HQ}

\begin{figure}
\includegraphics[width=\columnwidth]{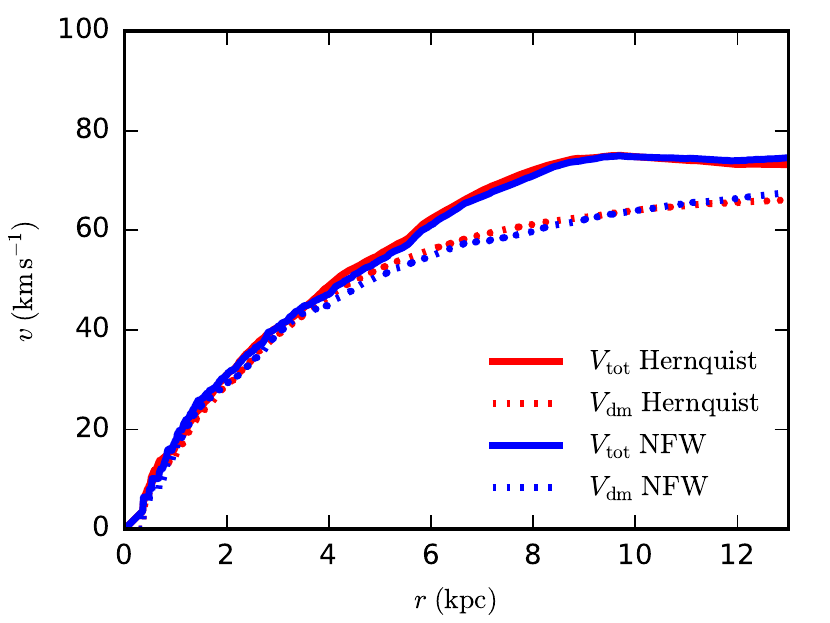}
\includegraphics[width=\columnwidth]{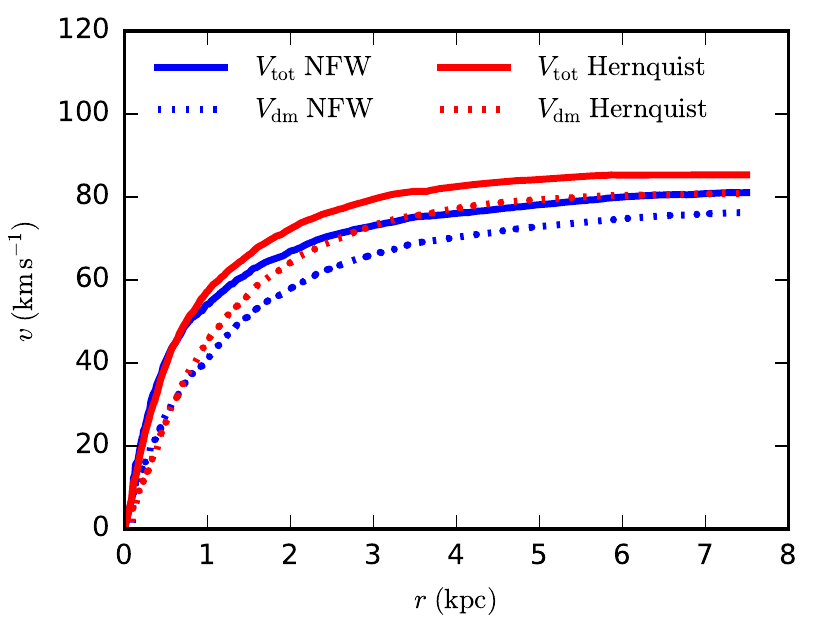}
\caption{Comparison of the evolution of `equivalent' Hernquist and NFW halos in the presence of baryonic disks. \emph{Upper panel} is a low concentration halo very similar to our IC~2574 fit, whilst \emph{lower panel} is high concentration setup similar to that of UGC~5721. The total rotation curves for the Hernquist halo are shown in \emph{solid red} vs. the NFW in \emph{solid blue}, with the dark matter contributions in \emph{dotted lines} respectively.}
\label{fig:HQ_vs_NFW}
\end{figure}

As mentioned in Section~\ref{sec:bar_ic}, the NFW and Hernquist initial conditions can evolve slightly differently in SIDM, driven by their velocity dispersion profiles (the NFW profile has somewhat increased velocity dispersion just beyond the scale radius $r_{\rm s}$). In particular once the radius of unit scattering $r_1$ has saturated $r_{\rm s}$, the Hernquist halo has slightly less `heat' (i.e. velocity dispersion) for SIDM to transfer than NFW. Since halos in cosmological N-body simulations have profiles closer to NFW than to Hernquist, we wish to test whether there are systematic differences due to the use of Hernquist halos for our SIDM simulations.

In order to test this, we performed simulations similar to those of the IC~2574 and UGC~5721 cases (although not actually our final best-fits given in Table~\ref{table:gal_params_fits}). We created NFW halos using the publicly available {\sc Spheric}\footnote{\url{https://bitbucket.org/migroch/spheric}} code \citep{GarrisonKimmel_2013} and equivalent Hernquist halos matched via Eqn.~(\ref{eq:rhq}). Notably the mass distributions are not the same outside the scale radius $r_{\rm s}$ even in the ICs, so we do not expect perfect fits even ignoring the differences in velocity dispersion profiles. In Fig.~\ref{fig:HQ_vs_NFW} we have plotted the profiles after 10~Gyr of evolution with $\som=3\;\cmspg$ with the the dark matter contribution and the total rotation curves implied by a realistic baryonic contribution. One can see that for the low concentration halo (upper panel), the differences are truly negligible. On the other hand, for the higher concentration halo (lower panel, very similar to the $+0.9\sigma$ of UGC~5721), the extra velocity dispersion (and consequently larger core radius) of the evolved NFW case becomes apparent, albeit only at the level of $\approx 10\%$. From this we infer that the central densities from high concentration halos are overestimated with the Hernquist model, and if one wishes to use equivalent NFW halos one should increase the concentration (use a shorter $r_{\rm s}$) by a few percentage points, including those parameters in Section~\ref{sec:obs_curve_sims}.

\end{document}